\documentclass[
twocolumn,
superscriptaddress,
amsmath,amssymb,
prl,
]{revtex4-2}
\usepackage{graphicx}% Include figure files
\usepackage{hyperref}
\usepackage{dcolumn}% Align table columns on decimal point
\usepackage{bm}% bold math
\usepackage{epstopdf}
\usepackage{romannum}
\usepackage{csquotes}
\usepackage[dvipsnames]{xcolor}
\usepackage{hyperref}% add hypertext capabilities
\usepackage{array}
\usepackage{booktabs}%表格线安装包
\begin{document}
\preprint{FeTeSe-Hc2}

\title{Novel Anisotropy of Upper Critical Fields in Fe$_{1+y}$Te$_{0.6}$Se$_{0.4}$}

\author{Yongqiang Pan}
\affiliation{Department of Physics, Southeast University, Nanjing 211189, China}
\affiliation{Key Laboratory of Materials Physics, Institute of Solid State Physics, Chinese Academy of Sciences, Hefei, 230031, China}

\author{Yue Sun}
\email{Corresponding author: sunyue@seu.edu.cn}
\affiliation{Department of Physics, Southeast University, Nanjing 211189, China}

\author{Nan Zhou}
\affiliation{Department of Physics, Southeast University, Nanjing 211189, China}
\affiliation{Key Laboratory of Materials Physics, Institute of Solid State Physics, Chinese Academy of Sciences, Hefei, 230031, China}

\author{Xiaolei Yi}
\affiliation{Department of Physics, Southeast University, Nanjing 211189, China}

\author{Jinhua Wang}
\affiliation{School of Electrical and Electronic Engineering, Huazhong University of Science and Technology, Wuhan 430074, China}

\author{Zengwei Zhu}
\affiliation{School of Electrical and Electronic Engineering, Huazhong University of Science and Technology, Wuhan 430074, China}

\author{Hiroyuki Mitamura}
\affiliation{Institute for Solid State Physics, The University of Tokyo, Kashiwa 277-8581, Japan}

\author{Masashi Tokunaga}
\affiliation{Institute for Solid State Physics, The University of Tokyo, Kashiwa 277-8581, Japan}

\author{Zhixiang Shi}
\email{Corresponding author:zxshi@seu.edu.cn}
\affiliation{Department of Physics, Southeast University, Nanjing 211189, China}

\date{\today}

\begin{abstract}
\textbf{}
Studying the upper critical field ($\mu_0$$H$$_{\rm{c2}}$) and its anisotropy of superconductors is of great importance because it can provide an unusual insight into the pair-breaking mechanism. Since Fe$_{1+y}$Te$_{1-x}$Se$_x$ exhibits the high $\mu_0$$H$$_{\rm{c2}}$ and small anisotropic superconductivity, it has attracted considerable attention. However, some issues related to $\mu_0$$H$$_{\rm{c2}}$ are still unknown, including the effect of excess Fe content on $\mu_0$$H$$_{\rm{c2}}$ behavior and the origin of the crossover of the $\mu_0H_{\rm{c2}}^c $ -- $ T$ and $\mu_0H_{\rm{c2}}^{ab}$ -- $T$ curves. In this work, the value of $\mu_0$$H$$_{\rm{c2}}$ of Fe$_{1+y}$Te$_{0.6}$Se$_{0.4}$ single crystals with controlled amounts of excess Fe was obtained by resistivity measurements over a wide range of temperatures down to $\sim$ 1.5 K, and magnetic fields up to $\sim$ 60 T. The crossover of the $\mu_0H_{\rm{c2}}^c $ -- $ T$ and $\mu_0H_{\rm{c2}}^{ab}$ -- $T$ curves was found to be independent of the excess Fe content. The angle dependence of $\mu_0H_{\rm{c2}}$ was also checked. The $\mu_0H_{\rm{c2}}(\theta)$ symmetry at higher temperature near $T_c$ could be fitted by anisotropic G-L model, and novel fourfold symmetry of $\mu_0H_{\rm{c2}}$ at lower temperature was found. Based on our spin-locking pairing model, the crossover behavior originates from the anisotropic spin-paramagnetic effect, and the novel fourfold symmetry of $\mu_0H_{\rm{c2}}$ could be understood by our extended anisotropic G-L model.
%\begin{description}
%\item[PACS numbers]
%\verb+74.70.Xa+, \verb+74.62.En+, \verb+74.25.fc+
%\end{description}
\end{abstract}
%\pacs{Valid PACS appear here}% PACS, the Physics and Astronomy
                             % Classification Scheme.
%\keywords{Suggested keywords}%Use showkeys class option if keyword
                              %display desired
\maketitle

\section{Introduction}
\textbf{}
The upper critical field $\mu_0$$H$$_{\rm{c2}}$ is sensitive to microscopic superconducting (SC) parameters (e.g., the SC energy gap $\triangle _{\rm{SC}}$ and the mean free path $\ell$)\cite{mercure2012upper,chen2008superconducting,braithwaite2010evidence} and is beneficial for understanding the unconventional superconductivity, including the coherence length $\xi$, electronic structure, and pair-breaking mechanism\cite{hunte2008two,zhuang2015pauli}. According to the Bardeen-Cooper-Schrieffer (BCS) theory, the superconductivity is based on Cooper pairs. Cooper pairs are made up of two electrons with opposite spins and momenta, which can be depaired by an external magnetic field via two primary mechanisms\cite{fuchs2009orbital}. One is the orbital depairing mechanism involving the Lorentz force (orbital depairing effect, ODE), which is dominant in the high-temperature region. The other is the Pauli spin-paramagnetic depairing mechanism involving the Zeeman effect (spin-paramagnetic depairing effect, SPDE)\cite{agosta2017calorimetric,lei2012iron}, which is dominant in the low-temperature region. For single-band superconductors, according to the Werthamer, Helfand, and Hohenberg (WHH) theory\cite{werthamer1966temperature}, the two primary mechanisms are expressed using two dimensionless parameters, the Maki parameter $\alpha$ and spin-orbit scattering $\lambda_{\rm{so}}$\cite{mercure2012upper,maki1966effect}. For multiband superconductors, $\mu_0$$H_{\rm{c2}}$ can be described successfully using the two-band BCS model\cite{baily2009pseudoisotropic,gurevich2003enhancement,hunte2008two,jaroszynski2008upper,lee2009effects,kano2009anisotropy,golubov2002specific,xing2017two} and two-band Ginzburg-Landau (G-L) theory\cite{grigorishin2016effective,askerzade2006ginzburg}. The two-band G-L theory yields a nonlinear temperature dependence of $\mu_0$$H$$_{\rm{c2}}$($T$) when the temperature is close to the critical temperature $T_{\rm{c}}$, which is different from the linear behavior expected in the single-band theory. Additionally, the anisotropy $\gamma$ of $\mu_0$$H$$_{\rm{c2}}$ ($\gamma_H$ = $H_{\rm{c2}}^{ab}$/$H$$_{\rm{c2}}^c$, where $H$$_{\rm{c2}}^{ab}$ and $H$$_{\rm{c2}}^c$ are the $H$$_{\rm{c2}}$ when $H$$\parallel$$ab$ and $H$$\parallel$$c$, respectively) is related to the dimensionality and topology of the electronic structure\cite{balakirev2015anisotropy}, which is crucial for understanding multiband effects.

Iron-based superconductors (IBSs) exhibit rich distinctive features, such as the two-band effect, ODE, and SPDE, which lead to a peculiar temperature dependence of $\mu_0$$H_{\rm{c2}}$($T$)\cite{lei2012iron,yuan2009nearly,zhang2011upper}. Among IBSs, the 11-system is unique in its structural simplicity which is favorable for probing the SC pairing mechanism. Recent reports have shown that Fe$_{1+y}$Te$_{1-x}$Se$_x$ presents a strong-coupling superconductivity, a strong electron correlation, and a crossover from BCS coupling to Bose-Einstein-condensation (BEC) coupling\cite{lubashevsky2012shallow,okazaki2013superconductivity,rinott2017tuning,wang2016upper}. For the upper critical fields, $\mu_0$$H$$_{\rm{c2}}^c$ of Fe$_{1+y}$Te$_{0.6}$Se$_{0.4}$ shows a multiband behavior without the SPDE, which can be fitted by the two-band model\cite{gurevich2003enhancement}. By contrast, $\mu_0$$H$$_{\rm{c2}}^{ab}$ shows a single-band behavior due to the SPDE. Besides, the impurity density, especially the excess Fe stoichiometry $y$\cite{sun2019review}, greatly affects the value of $\mu_0H_{\rm{c2}}$(0 K) and behavior of $\mu_0H_{\rm{c2}}(T)$. In the work by Matsuura $et$ $al$, $\mu_0H_{\rm{c2}}$ (0 K) depends on the impurity concentration, as does the initial slope of $\mu_0H_{\rm{c2}}(T)$, which will increase with increasing impurity content\cite{salamon2016upper,pan2021anisotropic,abrikosov1960contribution,matsuura1977theory,fuchs2009orbital,tinkham1996introduction,braithwaite2010evidence,klein2010thermodynamic,stoner1945xcvii}. However, Fe$_{1+y}$Te$_{0.6}$Se$_{0.4}$ samples with fewer impurities present higher $\mu_0H_{\rm{c2}}(T)$, which is abnormal and need to be studied.

\begin{figure*}
	\includegraphics[width=40pc]{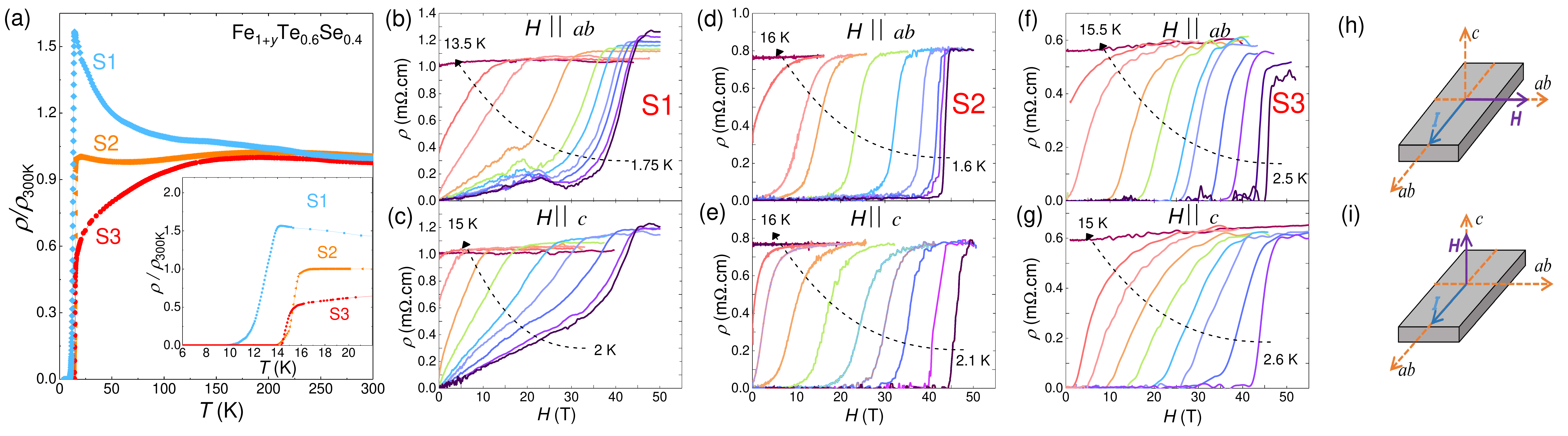}
	%\begin{center}
	\caption{\label{fig1} (a) Temperature dependence of the resistivity \textit{$\rho$} reduced by \textit{$\rho$}$_{\rm{300K}}$ under zero field for three samples. The inset shows the enlarged region near $T_{\rm{c}}$. (b)--(g)Magnetic field dependence of resistivity of samples S1, S2, and S3, respectively.  (h)--(i) Schematics of the applied magnetic field directions for the cases of \textit{$H$}$\parallel$\textit{$ab$} and \textit{$H$}$\parallel$\textit{$c$}, respectively. }
	%\end{center}
\end{figure*}

Interestingly, a crossover happens in $\mu_0$$H$$_{\rm{c2}}$ at low temperature ($T_{\rm{cr}}$), i.e., $H_{\rm{c2}}^{ab}$ $>$ $H_{\rm{c2}}^c$ in the high-temperature region, whereas $H_{\rm{c2}}^{ab}$ $<$ $H_{\rm{c2}}^c$ in the low-temperature region. This crossover on  $\mu_0$$H$$_{\rm{c2}}$ is unusual and can be found in Ba$_{1-x}$K$_x$Fe$_2$As$_2$\cite{yuan2009nearly}, $A$Cr$_3$As$_3$\cite{Liang2019Upper}, $A_2$Cr$_3$As$_3$ \cite{balakirev2015anisotropy,Cao2018Superconductivity} ($A$ represent alkali metal). It is interpreted as the orbital limiting effect persisted at all field angles, or Ising-like spin-singlet superconductivity. Until now, the origin of this crossover has not been understood, which might be due to a peculiar SC pairing mechanism. In addition, the angle dependence of $\mu_0H_{\rm{c2}}(\theta)$ near the crossover point will be an interesting issue, since the crossover in $\mu_0H_{\rm{c2}}$($T$) implies the change of symmetry of $\mu_0H_{\rm{c2}}$($\theta$)\cite{noji2010growth}. In this work, a series of Fe$_{1+y}$Te$_{0.6}$Se$_{0.4}$ single crystals were synthesized with different amount of excess Fe $y$ ranging from 0 to 0.14. The values of $\mu_0$\textit{$H$}$_{\rm{c2}}$ were obtained under an external magnetic field up to $\sim$60 T. The $\mu_0H_{\rm{c2}}$ anisotropy was investigated at the temperatures above and below $T_{\rm{cr}}$ on the clean crystal free from excess Fe. A spin-locking pairing model has been proposed to explain the novel $\mu_0H_{\rm{c2}}$ anisotropy.

\section{Experimental Details}
Fe$_{1+y}$Te$_{0.6}$Se$_{0.4}$ single crystals were synthesized via the standard self-flux method as shown in our previous reports\cite{sun2013bulk,sun2012effects,sun2014dynamics}. The crystals with different amounts of excess Fe were prepared by annealing\cite{sun2019review}. The determination of the amount of excess Fe can be seen in our previous report\cite{sun2021comparative,supplement} labeled as S1 (as-grown sample, Fe$_{1.14}$Te$_{0.6}$Se$_{0.4}$), S2 (half-annealed sample, Fe$_{1.07}$Te$_{0.6}$Se$_{0.4}$), and S3 (fully-annealed sample, Fe$_{1.0}$Te$_{0.6}$Se$_{0.4}$)\cite{sun2021comparative}. The electrical transport measurements were performed using a commercial PPMS-9 (Quantum Design, $\sim$9 T) and a high magnetic field generated from a 60-T magnet with a pulse width of $\sim$70 ms at the Wuhan National High Magnetic Field Center (WHHMF) and with a pulse width of $\sim$36 ms at the Institute for Solid State Physics, The University of Tokyo\cite{mitamura2020improved}.

\section{Results and discussions}
Figure 1(a) shows the reduced zero-field resistivity $\rho$/$\rho$$_{\rm{300K}}$ (where $\rho$$_{\rm{300K}}$ is the resistivity at \textit{$T$} = 300 K) for the synthesized single crystals. $T$$_{\rm{c}}$ determined by the 50$\%$ normal-state resistivity are $\sim$13.3 K (sample S1), $\sim$14.6 K (sample S2), and $\sim$15.2 K (sample S3), respectively.  The residual resistivity ratio $RRR$, defined as $\rho_{\rm{300 K}}$/$\rho (T_{\rm{c}}^{\rm{onset}})$, are estimated as $\sim$0.74, $\sim$0.92, and $\sim$2 for S1, S2, and S3, respectively. The increase of $T_{\rm{c}}$ and decrease of residual resistivity manifest the improvement of sample quality after removing excess Fe\cite{sun2021comparative,sun2014multiband}. Magnetic field dependence of resistivity of S1, S2, and S3 are plotted at in Figs. 1(b)-(g). Temperature dependence of \textit{$\rho$}/\textit{$\rho$}$_{\rm{300K}}$ ($\rho$--$T$ curves) under different magnetic fields is shown in Fig. S1 (Supplement Materials)\cite{supplement}. The temperature dependence of $\mu_0$$H$$_{\rm{c2}}$ can be obtained from the $\rho$--$T$ and $\rho$--$H$ curves using the criterion of the 50$\%$ normal-state resistivity. With this criterion, the effects of the vortex motion expected from the 10$\%$ criterion and the SC fluctuation expected from the 90$\%$ criterion can be minimized\cite{xing2017two}. Additionally, $\mu$$_0$\textit{$H$}$_{\rm{c2}}$ obtained from both field sweeps (performed at the WHHMF) and temperature sweeps (using the PPMS) overlap with each other, demonstrating the consistence of the obtained $\mu_0$$H$$_{\rm{c2}}$. Reduced temperature dependence of $\mu$$_0$$H$$_{\rm{c2}}^{c}$($T$) and $\mu$$_0$$H$$_{\rm{c2}}^{ab}$($T$) for the three samples are shown in Figs. 2(a)--(c) by points. The red curves and dark cyan dash curves present the WHH fit on $\mu$$_0$$H$$_{\rm{c2}}^{ab}$($T$) and two band fit on $\mu$$_0$$H$$_{\rm{c2}}^{c}$($T$), respectively.

For $\mu$$_0$$H$$_{\rm{c2}}^{c}$($T$), the fitting using the single-band WHH model was not successful because of the linear $\mu$$_0$$H$$_{\rm{c2}}^{c}$ behavior at low temperatures \cite{supplement}, which is related to the multigap nature of Fe$_{1+y}$Te$_{0.6}$Se$_{0.4}$. Therefore, the two-band model was adopted, in the following form in the dirty limit\cite{gurevich2003enhancement}:

\begin{eqnarray}
	&& 0=a_0(\ln t+U(h)(\ln t+U(\eta h))\nonumber\\
	&& +a_1(\ln t+U(h))+a_2(\ln t+U(\eta h)),
\end{eqnarray}
where \textit{$a$}$_0$ = 2($\lambda$$_{11}$$\lambda$$_{22}$-$\lambda$$_{12}$$\lambda$$_{21}$)/$\lambda$$_0$, \textit{$a$}$_1$ = 1+($\lambda$$_{11}$-$\lambda$$_{22}$)/$\lambda$$_0$, \textit{$a$}$_2$ = 1-($\lambda$$_{11}$-$\lambda$$_{22}$)/($\lambda$$_0$/2), $\lambda$$_0$ = (($\lambda$$_{11}$-$\lambda$$_{22}$)$^2$+4$\lambda$$_{12}$$\lambda$$_{21}$)$^{1/2}$, \textit{$t$} = \textit{$T$}/\textit{$T$}$_{\rm{c}}$, \textit{$h$} = $\mu_0$\textit{$H$}$_{\rm{c2}}^{c}$\textit{$D$}$_1$/(2$\Phi$$_0$/\textit{$T$}), $\eta$ = \textit{$D$}$_2$/\textit{$D$}$_1$, and \textit{$U$}(x) = $\Psi$(1/2+x)-$\Psi$(1/2). $\Psi$(x) is the digamma function. \textit{$D$}$_1$ and \textit{$D$}$_2$ are the diffusivity of each band. $\lambda$$_{11}$ and $\lambda$$_{22}$ denote the intraband coupling constants which can be derived from the $\mu$SR experiment and can be adjusted\cite{khasanov2010evolution}. $\lambda$$_{12}$ and $\lambda$$_{21}$ are the interband coupling constants. Here, it is assumed that the intraband coupling dominates the $\mu$$_0$\textit{$H$}$_{\rm{c2}}^{c}$(\textit{$T$}), and the interband coupling takes the value $\lambda$$_{12}$ = $\lambda$$_{21}$ to reduce the number of free parameters. The fitting results of the $\mu$$_0$\textit{$H$}$_{\rm{c2}}^{c}$(\textit{$T$}) data obtained using the two-band model are shown in Figs. 2(a)-(c) by blue dash curves. The fitting parameters can be seen in Table I. We can find $\lambda$$_{12}$$\lambda$$_{21}$ $\ll$ $\lambda$$_{11}$$\lambda$$_{22}$, which is similar to MgB$_2$\cite{golubov2002specific}, indicating that the interband coupling is weak. The upper critical fields obtained from the two-band model ($\mu$$_0$$H$$_{\rm{c2}}^{c,\rm{TB}}$(0 K)) are 45.50 (S1), 50.88 (S2), and 52.20 T (S3). The corresponding $\xi_{ab}$(0 K), defined by the G-L equation $\xi$$_{ab}$(0 K) = ($\Phi$$_0$/2$\pi$$\mu_0$\textit{$H$}$^c_{c2}$(0 K))$^{1/2}$, are 2.69 (S1), 2.54 (S2), and 2.51 nm (S3), respectively. The slight increase of $\mu$$_0$$H$$_{\rm{c2}}^{c,\rm{TB}}$(0 K) and decrease of $\xi$$_{ab}$(0 K) with reducing of excess Fe are related to the increase of $T_{\rm{c}}$. Meanwhile, the values of -$\mu_0$d$H$$_{\rm{c2}}^c$/d$T$ at $T$$_{\rm{c}}$ are 3.64 (S1), 4.76 (S2), and 6.74 T/K (S3). The increase of -$\mu_0$d$H$$_{\rm{c2}}^c$/d$T$ at $T$$_{\rm{c}}$ may lead to the change of coupling strength from BCS coupling to BEC coupling\cite{gurevich2003enhancement,yuan2018universal,gurevich2011iron,supplement}

\begin{table}
	\centering
	\caption{Parameters for Fe$_{1+y}$Te$_{0.6}$Se$_{0.4}$.}
	\begin{tabular*}{0.5\textwidth}{@{\extracolsep{\fill}}ccccccccccccc}
		\hline
		\hline
		\specialrule{0em}{1pt}{1pt}%specialrule 命令第一个大括号控制表格线的粗细，若为0，则表格线透明，第二个大括号是表格线与上方内容的距离，第三个大括号是表格线与下方内容的距离，通过改变后两个大括号中的值来控制行高！
		& samples & S1 & S2 & S3 & \\
		\hline
		\specialrule{0em}{1pt}{1pt}
		& \textit{$RRR$} & 0.74 & 0.92 & 2 & \\
		\hline
		\specialrule{0em}{1pt}{1pt}
		& \textit{$T$}{$_{\rm{c}}$} (K) &13.2&14.6&15.2& \\
		\hline
		\specialrule{0em}{1pt}{1pt}
		& -$\mu_0$d$H_{\rm{c2}}^c$/d$T$$\mid$$_{T=Tc}$ (T/K) &3.64&4.76&6.74& \\
		\hline
		\specialrule{0em}{1pt}{1pt}
		& $\mu_0$\textit{$H$}{$_{\rm{c2}}^{c,\rm{TB}}$}(0 K) (T) &45.50&50.88&52.20& \\
		\hline
		\specialrule{0em}{1pt}{1pt}
		& $\xi$$_{ab}$(0 K) (nm) &2.69&2.54&2.51& \\
		\hline
		\specialrule{0em}{1pt}{1pt}
		& $\lambda$$_{11}$ &0.250&0.278&0.307& \\
		\hline
		\specialrule{0em}{1pt}{1pt}
		& $\lambda$$_{12}$($\lambda$$_{21}$) &0.050&0.006&0.001& \\
		\hline
		\specialrule{0em}{1pt}{1pt}
		& $\lambda$$_{22}$ &0.320&0.244&0.274& \\
		\hline
		\specialrule{0em}{1pt}{1pt}
		& $D_1$ &0.750&0.258&0.256& \\
		\hline
		\specialrule{0em}{1pt}{1pt}
		& $\eta$ &0.237&0.545&0.600& \\
		\hline
		\hline
		\specialrule{0em}{1pt}{1pt}
		& -$\mu_0$d\textit{$H$}{$_{\rm{c2}}^{ab}$}/d\textit{$T$}$\mid$$_T$$_=$$_T${$_{\rm{c}}$} (T/K) & 6.289 & 10.722 & 11.716 & \\
		\hline
		\specialrule{0em}{1pt}{1pt}
		& $\mu_0$\textit{$H$}{$_{\rm{c2}}^{ab,\rm{orb}}$}(0 K) (T) & 56.5 & 109.4 & 122.0 & \\
		\hline
		\specialrule{0em}{1pt}{1pt}
		& $\mu_0$\textit{$H$}{$_{\rm{c2}}^{ab}$}(0 K) (T) & 41.7 & 44.0 & 44.5 & \\
		\hline
		\specialrule{0em}{1pt}{1pt}
		& Maki parameter $\alpha$ & 1.30 & 3.45 & 3.90 & \\
		\hline
		\hline
	\end{tabular*}
\end{table}

\begin{figure}
	\includegraphics[width=16pc]{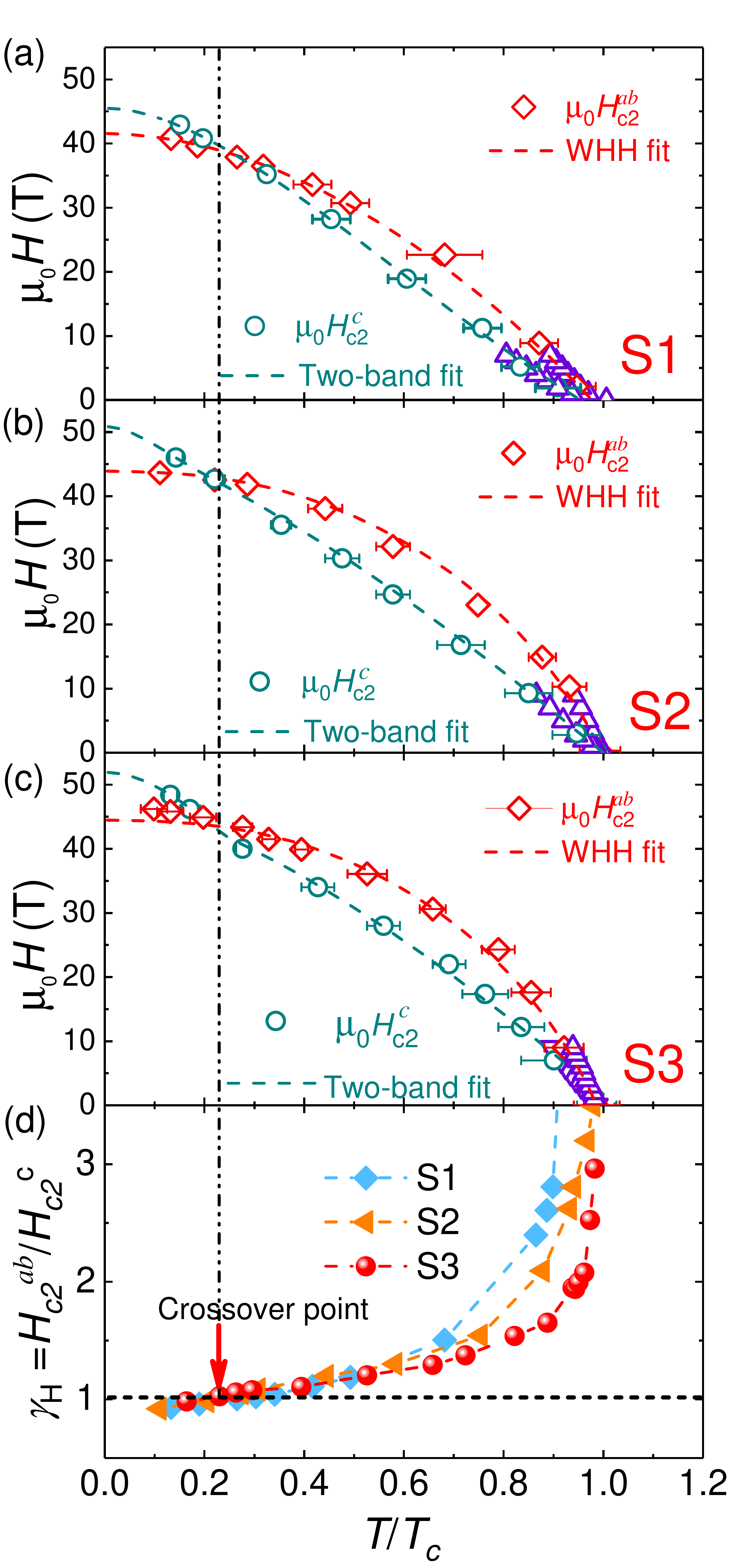}
	\begin{center}
		\caption{\label{fig2}Reduced temperature dependence of $\mu$$_0$$H$$_{\rm{c2}}^{c}$ and $\mu$$_0$\textit{$H$}$_{\rm{c2}}^{ab}$ of samples (a) S1, (b) S2, and (c) S3. Hollow diamond and circle represent the $\mu_0H_{\rm{c2}}$ obtained from HMF. Hollow triangle represent the $\mu_0H_{\rm{c2}}$ obtained from PPMS. (d) Anisotropic $\gamma_H$ of samples S1, S2 and S3. The red arrow indicates the crossover points of $\mu$$_0$$H$$_{\rm{c2}}$. As the black dash line shown, all crossover are located at same $T/T_{\rm{c}}$.}
	\end{center}
\end{figure}

For $\mu_0$$H$$_{\rm{c2}}^{ab}$($T$), these $\mu_0$$H$$_{\rm{c2}}^{ab}$($T$) curves show a similar convex shape. The $\mu_0$$H$$_{\rm{c2}}^{ab}$(0 K) of all samples exceed 40 T. The WHH model considering both the Maki parameter $\alpha$ and the spin-orbital effect parameter $\lambda$$_{\rm{so}}$ is used to fit $\mu_0$$H$$_{\rm{c2}}^{ab}$($T$)\cite{khim2010evidence}. As shown in Figs. 2(a)--(c) and Table I, the orbital field $\mu$$_0$$H$$_{\rm{c2}}^{ab,\rm{orb}}$(0 K) for three samples defined by $-$0.693$T$$_{\rm{c}}$$\mu_0$d$H$$_{\rm{c2}}$/d$T$$\mid$$_T$$_=$$_T$$_{\rm{c}}$ is 56.5 (S1), 109.4 (S2), and 122.0 T (S3), these values are larger than FeSe\cite{zhou2021disorder} but smaller than the typical values of 122 system\cite{tarantini2011significant,gasparov2011electron} and 112 system\cite{xing2017two}. In order to investigate the influence of impurities on $\mu_0H_{\rm{c2}}^{ab}$, several important parameters and their variation tendency are noticed. For sample S1, a small $\alpha$ ($\sim$1.3) results in its $\mu$$_0$$H$$_{\rm{c2}}^{ab}$(0 K) ($\sim$ 41.7 T) being not particularly small, meaning that the SPDE is comparatively weak. Contrarily, for sample S3, a large $\alpha$ $\sim$3.9 results in a limited $\mu$$_0$$H$$_{\rm{c2}}^{ab}$(0 K) ($\sim$44.5 T), which indicates that the SPDE is enhanced by removing excess Fe.  The $RRR$ dependence of $\mu$$_0$\textit{$H$}$_{\rm{c2}}^{ab}$(0 K) and the Maki parameter $\alpha$ are plotted in Fig. 3 (orange and green areas). These parameters exhibit similar increasing behavior with increasing $RRR$. We speculated that the disorder, i.e., excess Fe, not only suppresses the $T_c$, but also weakens the SPDE and reduces $\mu_0$$H$$_{\rm{c2}}^{ab}$. Meanwhile, the coherence length along $c$-axis of S3 is also calculated to be 9.5 {\AA} by two-band BCS model ($\xi_{c}^{\rm{BCS}}$(0 K)) \cite{supplement}. This value is larger than previous report $\sim$4.4 \AA \cite{gurevich2011iron}. Compared with the lattice parameter $c$ $\sim$ 6 \AA, this small $\xi_{c}^{\rm{BCS}}$(0 K) indicates that Fe$_{1+y}$Te$_{0.6}$Se$_{0.4}$ may show some quasi-two-dimensional behavior. Furthermore, $\mu$$_0$$H$$_{\rm{c2}}^{ab}$ is smaller than $\mu$$_0$$H$$_{\rm{c2}}^{ab,\rm{orb}}$, and the difference becomes larger at low temperatures, indicating that the SPDE is dominant at the low-temperature region.

\begin{figure}
	\includegraphics[width=20pc]{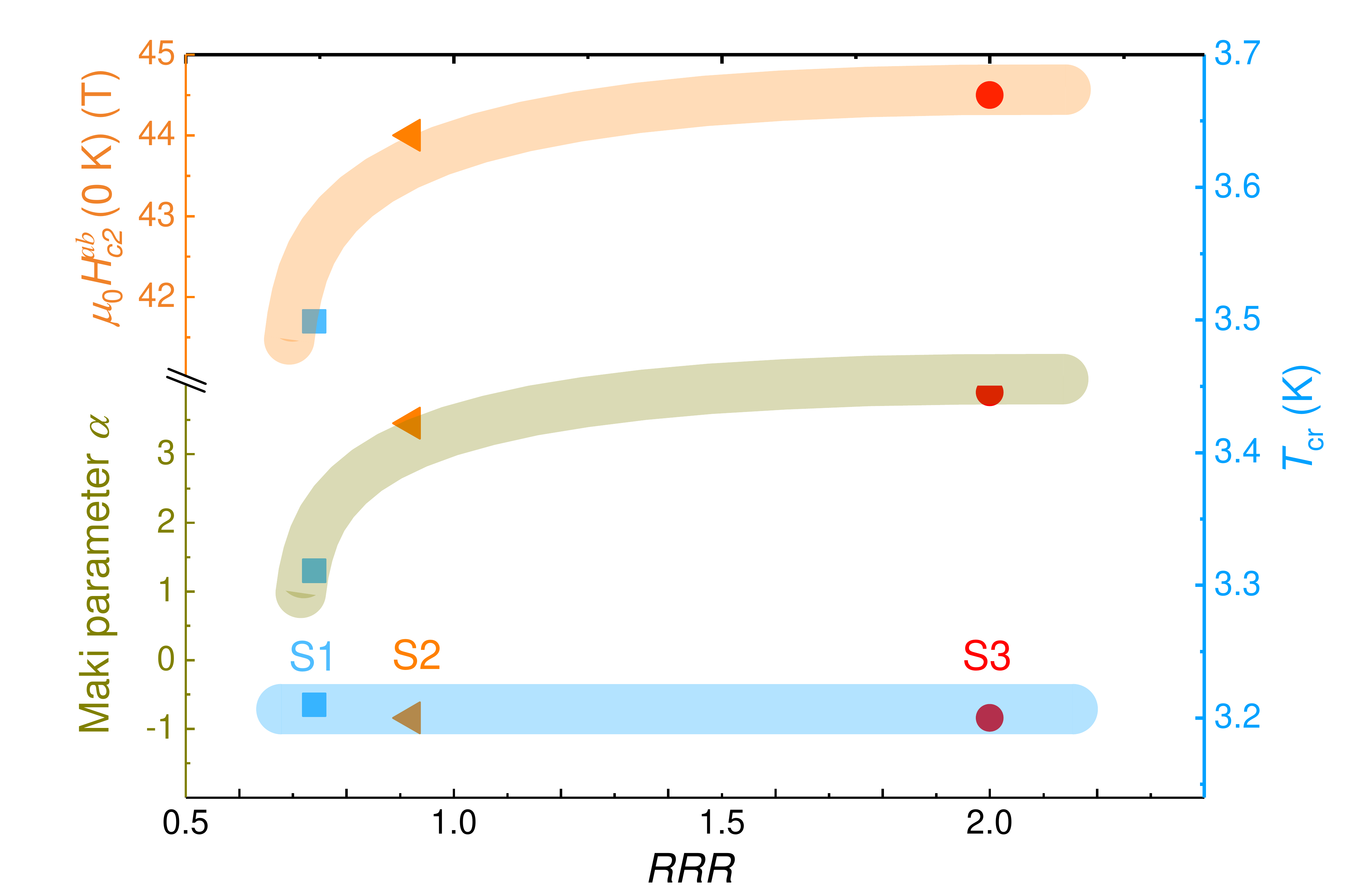}
	\begin{center}
		\caption{\label{fig3}\textit{$RRR$} dependencies of $\mu$$_0$\textit{$H$}$_{\rm{c2}}^{ab}$(0 K), the Maki parameter $\alpha$, and \textit{$T$}$_{\rm{cr}}$ of samples S1, S2, and S3, respectively. Those colored areas are represent the variation tendency of parameters, its are corresponding to colored vertical coordinates.}
	\end{center}
\end{figure}

The crossover can be observed in Figs. 2(a)-(d). Figure 2(d) shows the temperature dependence of the $\mu$$_0H_{\rm{c2}}$ anisotropy $\gamma_H$($T$) defined as $\mu_0$$H$$_{\rm{c2}}^{ab}$/$\mu_0$$H$$_{\rm{c2}}^{c}$. As the temperature decreases, $\gamma_H$($T$) finally drops below 1 after the crossover temperature $T_{\rm{cr}}$ (the red arrow in the Fig. 2(d)). Interestingly, the $T_{\rm{cr}} /T_c$ are unchanged in different Fe$_{1+y}$Te$_{0.6}$Se$_{0.4}$ samples (shown as the black dashed line, $T_{\rm{cr}}$ = 0.22$T_{\rm{c}}$ $\sim$3.2 K). The value of $T_{\rm{cr}}$ is plotted in Fig. 3 (blue area), it indicates that the crossover is disorder-robust and intrinsic property. The crossover of $\mu_0$$H$$_{\rm{c2}}$ in two directions is a novel phenomenon observed in Fe(Te,Se) and (Ba,K)Fe$_2$As$_2$\cite{baily2009pseudoisotropic,yuan2009nearly,lei2012iron}, and its origin is still unclear. Previous works have tried to explain this crossover in terms of the strong SPDE, that suppresses the $\mu$$_0$\textit{$H$}$_{\rm{c2}}^{ab}$ at low temperatures\cite{lei2012iron}. However, the origin of such a strongly anisotropic SPDE remains unknown.

\begin{figure*}
	\includegraphics[width=42pc]{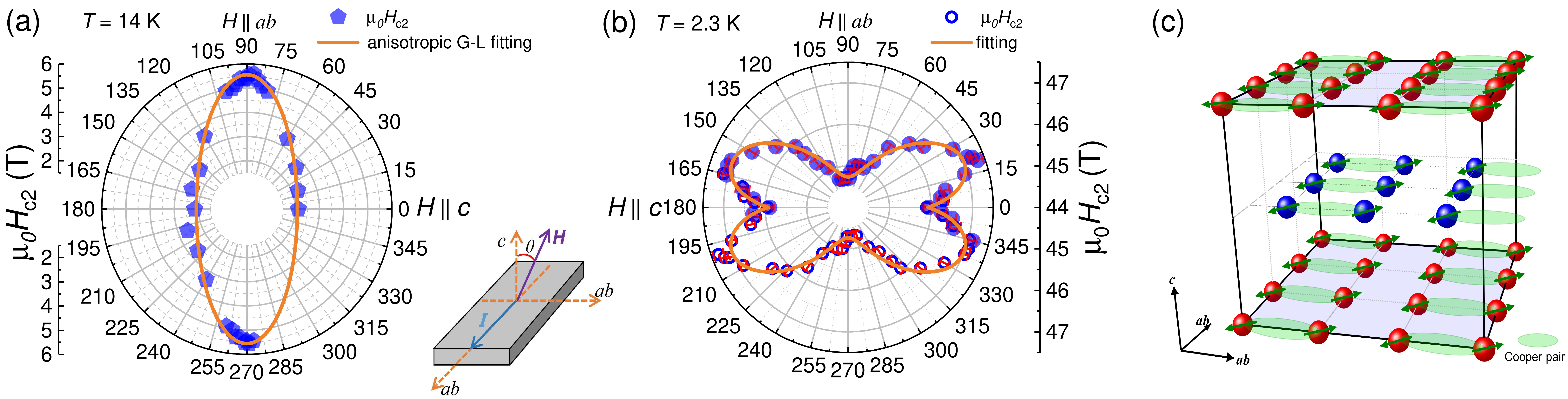}
	%\begin{center}
	\caption{\label{fig4}(a) Angle $\theta$ dependent $\mu_0H_{\rm{c2}}(\theta)$ at $T$ = 14 K, which are shown by blue solid points. The orange curve represents the anisotropic G-L model fitting result. Inset is a schematic of the applied magnetic field directions. (b) Angle $\theta$ dependent $\mu_0H_{\rm{c2}}(\theta)$ in the range of -15$^\circ$ $\sim$ 200$^\circ$ at $T$ = 2.3 K, which are shown by blue solid points. The hollow points represent the 180$^\circ$ rotated points of $\mu_0H_{\rm{c2}}(\theta)$ in the range of -15$^\circ$ $\sim$ 200$^\circ$. The orange curve represents the extended anisotropic G-L model fitting result. (c) Schematic diagram of the spin-locking phenomenon of Cooper pairs. The spheres represent carriers and the arrows represent their spin direction. The pair of arrows in different layers is either parallel or antiparallel. The green ellipse signify the Cooper pair is constitutive of two carriers.}
	%\end{center}
\end{figure*}

The appearance of crossover implies a dramatic change on $\mu_0H_{\rm{c2}}$ anisotropy. In order to investigate the symmetry of $\mu_0H_{\rm{c2}}$ above and below $T_{\rm{cr}}$, we further measured the angle $\theta$ ($\theta$ is the angle between $H$ and $c$-axis) dependence of magneto-resistance of S3 sample free from excess Fe at 14 K (near $T_{\rm{c}}$) and 2.3 K (below $T_{\rm{cr}} \sim$ 3.2 K), respectively. The angle $\theta$ dependence of $\mu_0H_{\rm{c2}}(\theta)$ was obtained according to the magneto-resistance data (The raw data of $R(H)$ in different field directions are shown in Fig. S4 in supplement materials). $\mu_0H_{\rm{c2}}(\theta)$ at 14 K is shown in Fig. 4(a), exhibiting a ellipse shape with twofold symmetry. The maximum value of $\mu_0H_{\rm{c2}}(\theta)$ appears at $H||ab$. The anisotropic G-L model\cite{xing2016anisotropic,blatter1992isotropic,tinkham1996introduction,thoutam2015temperature} was used to fit the $\mu_0H_{\rm{c2}}(\theta)$ at 14 K,
\begin{eqnarray}
	&&H_{\rm{c2}} ^2(\theta)=\frac{(H_{\rm{c2}}^{ab})^2}{\rm{sin}^2 \theta + \gamma_{GL} ^2 \rm{cos} ^2 \theta} .
\end{eqnarray}
The anisotropy parameter $\gamma_{\rm{GL}}$ estimated from the anisotropic G-L model fitting is 2.24, consistent with the $\gamma_{H}$  $\sim$2.4 estimated from $R$--$T$ curves under fields ($\mu_0H_{\rm{c2}}^{ab}$(14 K)$/\mu_0H_{\rm{c2}}^c$(14 K)).

The $\mu_0H_{\rm{c2}}(\theta)$ at 2.3 K are shown in Fig. 4(b). At this temperature, $\mu_0H_{\rm{c2}}^c$ is slightly larger than $\mu_0H_{\rm{c2}}^{ab}$. $\mu_0H_{\rm{c2}}(\theta)$ at 2.3 K shows a butterfly-pattern with fourfold symmetry. Two minimum values appear in the directions of $H||ab$ and $H\|c$, suggesting the SPDE get its largest value when $H\|ab$, while the ODE get its largest value when $H\|c$ (shown as the arrows in Fig. 4(b)). In theory, SPDE is interpreted as depairing mechanism due to the Zeeman effect aligning the spins of the two electrons with the applied field paralleled to $ab$-plane, while ODE is interpreted as depairing mechanism due to the Lorentz force acting via the charge on the momenta of the paired electrons with the applied field parallel to $c$-axis\cite{fuchs2009orbital,mockli2020ising}. Besides, two maxima appear near 22$^\circ$ (338$^\circ$) and 158$^\circ$ (202$^\circ$), suggesting that both ODE and SPDE are angle dependent, and ODE's attenuation rate exceeds the enhancement rate of in-plane SPDE when the $H$ rotate from 0$^\circ$ ($H\|c$) to 90$^\circ$ ($H\|ab$).

To explain the observed strongly anisotropic SPDE and novel $\mu_0H_{\rm{c2}}^c$($\theta$) anisotropy, a spin-locking pairing model in quasi-two-dimensional superconductors is proposed\cite{xing2017two} . In this model, the half-itinerant carriers are proposed. In a normal state with $T$ slightly above the pre-pairing critical temperature $T_{\rm{pf}}$, the majority of carriers are half-itinerant, i.e., their spin orientation is locked in the ${ab}$-plane but their charge and  spin are itinerant. These half-itinerant carriers are mainly from Fe ions in Fe$_{1+y}$Te$_{1-x}$Se$_x$ lattice, as shown in the schematic in Fig. 4(c). When $T$ $\leq$ $T_{\rm{c}}$, these half-itinerant carriers constitute Cooper pairs, and the long-range phase coherence of these Cooper pairs and supercurrent are formed. The spin-locked Cooper pairs display a peculiar anisotropy under an applied magnetic field. For $H$ $\parallel$ $ab$-plane, an angle exists between the spin direction of two carries in a Cooper pair and $H$, their magnetization energy are +$\mid$$\vec {M}$$\cdot$$\vec {H}$$\mid$ and -$\mid$$\vec {M}$$\cdot$$\vec {H}$$\mid$, and the depairing energy reaches a maximum value, which causes the paramagnetic effect to be dominant for $\mu_0$$H_{\rm{c2}}^{ab}$. When $H$ $\parallel$ $c$-axis ($H$ is perpendicular to the spin of all carriers of Cooper pairs), the magnetization energy is always zero, and no SPDE occurs. In this case, the ODE is dominant for $\mu_0$$H_{\rm{c2}}^{c}$, and the two-band effect could uncover in Fe$_{1+y}$Te$_{0.6}$Se$_{0.4}$. This model provides a good explanation of the origin of the anisotropic SPDE.

Considering the spin-locking pairing model, an extended anisotropic G-L model was used to fit the $\mu_0H_{\rm{c2}}(\theta)$ at 2.3 K. Due to spin-locked Cooper pairs, the angle-dependent Zeeman splitting energy is assumed as a separate term described by $\Delta E_Z = 2|\vec{M}\cdot\vec{H}| =|2g_{ab}\mu\sin\theta)|$ \cite{khim2021field,wang2020flux,kuchinskii2017temperature,sun2017effect}. The ODE ($\Delta E_{orb}$) could be expressed with the formation of  the Lorenz force $F_L$, $\Delta E_{orb} = |2F_L\cdot\xi_{GL}(\theta) /2| = |eHv_{ab}|\xi_{\rm{GL}}(\theta)$\cite{fuchs2009orbital}. Here, the $g_{ab}$ and $v_{ab}$ are  the Lande factor and Fermi velocity in $ab$-plane, respectively (The derivation of the formula can be found in  supplement materials). Taking both Zeeman splitting effect and ODE in anisotropic G-L model to get the extended anisotropic G-L model\cite{supplement}, the $\mu_0H_{\rm{c2}}(\theta)$ can be written as:
\begin{widetext}
	\begin{eqnarray}
		&& H_{c2}(\theta)= \frac{\hbar ^2} {2m^*\frac{(\xi_{GL}^{ab})^2}{\sin^2 \theta +\gamma_{GL}^2 \cos ^2 \theta}[\frac{\hbar e}{m^*c} +|2g_{ab}\mu_B \sin \theta| +|ev_{ab}| (\frac{(\xi_{GL}^{ab})^2}{\sin^2 \theta+\gamma_{GL}^2 \cos ^2 \theta})^{0.5}]} ,
	\end{eqnarray}
\end{widetext}
where $\mu_{\rm{B}}$, $m^*$, and $\hbar$ are Bohr magneton, electron effective mass, and reduced Planck constant, respectively. The fitting result is shown in Fig. 4(b) by orange curve. Two minimum values appear when angle $\theta = 0^\circ$, $90^\circ$, $180^\circ$, and $270^\circ$, indicates that both SPDE and ODE have strong directionality, which is consistent with the spin-locking pairing model. $\xi_{\rm{GL}}^{ab}$(2.3 K) and anisotropy parameter $\gamma_{\rm{GL}}$ of $\mu_0H_{\rm{c2}}$(2.3 K) obtained from the extended anisotropic G-L model are 2.73 nm and 1.24, respectively, which is consistent with those shown in Table I. Theoretically, the SPDE in the $ab$-plane is enhanced quickly with decreasing temperature, resulting in an obvious suppression on $\mu_0H_{\rm{c2}}^{ab}$ at low temperatures. On the other hand, the ODE along $c$-axis is enhanced slowly with decreasing temperature, leading the value of $\mu_0H_{\rm{c2}}^c$ to linearly increase and overshoot $\mu_0H_{\rm{c2}}^{ab}$ below $T_{\rm{cr}}$. When $H$ rotate away from 90$^\circ$ (0$^\circ$), SPDE (ODE) weaken quickly, resulting in a larger $\mu_0H_{\rm{c2}}(\theta)$ than it at $ab$-plane ($c$-axis). Therefore, the $\mu_0H_{\rm{c2}}(\theta)$ changes from a twofold symmetry near $T_{\rm{c}}$ to a fourfold symmetry at low temperatures.

\section{Conclusion}
$\mu_0$$H$$_{\rm{c2}}$ of Fe$_{1+y}$Te$_{0.6}$Se$_{0.4}$ single crystals with selected amounts of excess Fe were investigated by conducting resistivity measurements over a wide range of temperatures and magnetic fields. $\mu$$_0$$H$$_{\rm{c2}}^{c}$ and $\mu$$_0$$H$$_{\rm{c2}}^{ab}$ were fitted by the two-band model and the WHH model, respectively. The crossover observed on $\mu$$_0$$H$$_{\rm{c2}}$ is disorder-robust and indicates the presence of a strong anisotropic SPDE. Furthermore, the angle dependent $\mu$$_0$$H$$_{\rm{c2}}(\theta)$ exhibits a novel anisotropy with a twofold symmetry near $T_c$, but a fourfold symmetry at low temperatures. To understand the strong anisotropic SPDE and the novel anisotropy of $\mu$$_0$$H$$_{\rm{c2}}$, a spin-locking model was proposed and the novel fourfold symmetry of $\mu$$_0$$H$$_{\rm{c2}}(\theta)$ could be fitted by our extended anisotropic G-L model successfully.

\section*{Acknowledgments}
The present work was partly supported by the National Key R$\&$D Program of China (Grant No. 2018YFA0704300), the Strategic Priority Research Program of Chinese Academy of Sciences (Grant No. XDB25000000), and the National Natural Science Foundation of China (Grant No. U1932217, No. 12204487).

Yongqiang Pan, Yue Sun, and Nan Zhou contributed equally to this paper.

\acknowledgements

\bibliographystyle{unsrt}%
\bibliography{ref}% Produces the bibliography via BibTeX.

\pagebreak
\clearpage%强制分离浮动体
\newpage

\onecolumngrid
\begin{center}
	\textbf{\huge Supplemental information}
\end{center}
\vspace{1cm}
\twocolumngrid

\setcounter{equation}{0}
\setcounter{figure}{0}
\setcounter{table}{0}

\makeatletter
\renewcommand{\theequation}{S\arabic{equation}}
\renewcommand{\thefigure}{S\arabic{figure}}

\section{Sample Characterizations}
Temperature dependence and magnetic field dependence of resistivity of S1, S2, and S3 samples have been plotted in Fig. S1. Removing excess Fe can improve the superconductivity of Fe$_{1+y}$Te$_{0.6}$Se$_{0.4}$\cite{sun2021comparative,sun2019review,fang2010weak,sun2016influence}.

\begin{figure*}
	\includegraphics[width=38pc]{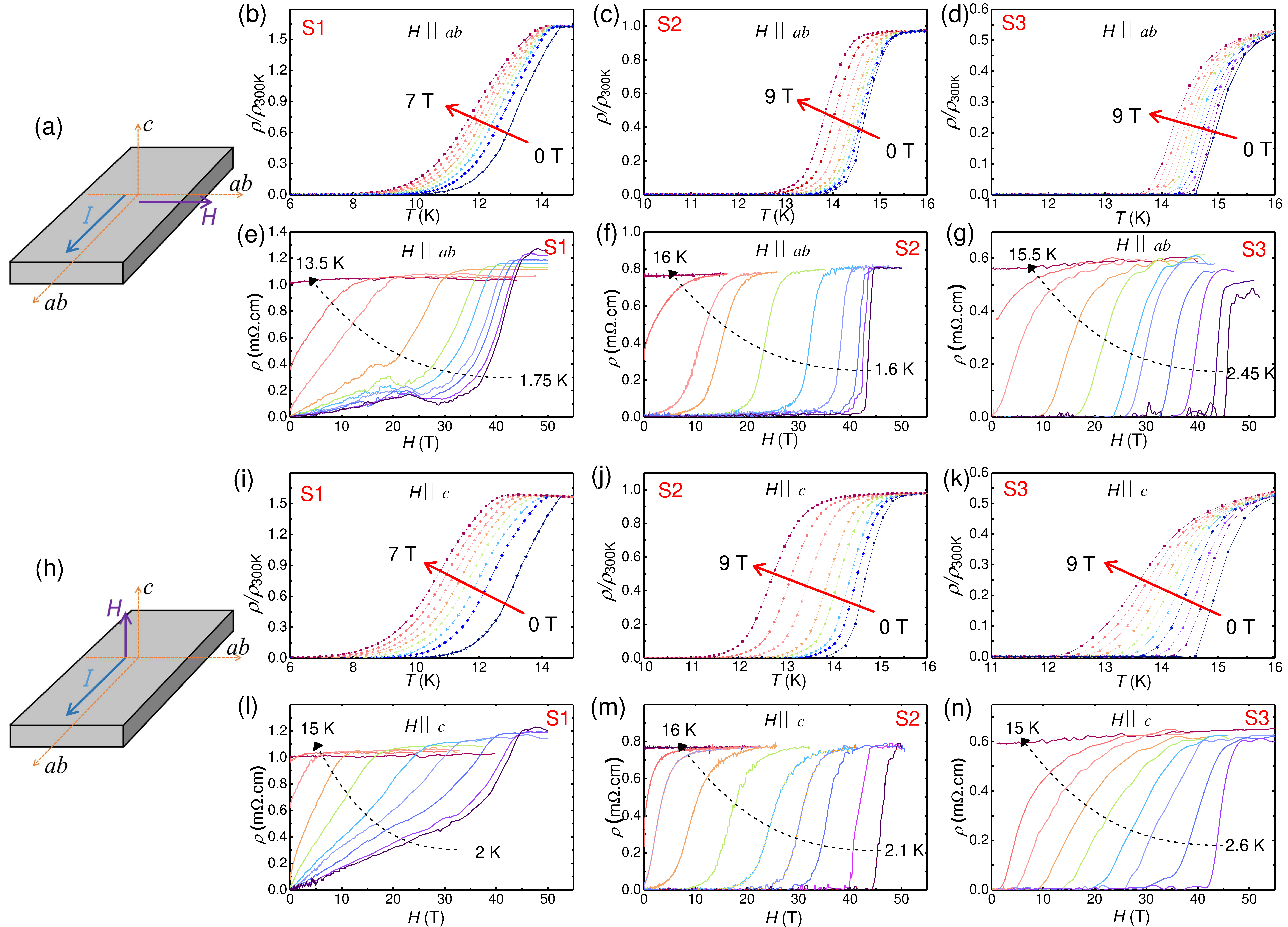}
	%\begin{center}
	\caption{\label{fig1}(a) and (h) Schematics of the applied magnetic field directions for the cases of \textit{$H$}$\parallel$\textit{$ab$} and \textit{$H$}$\parallel$\textit{$c$}, respectively. (b)--(d) Temperature dependence of the resistivity of samples S1 (under $H$ = 0, 1, 2, 3, 4, 5, 6, and 7 T), S2 (under $H$ = 0, 0.5, 1, 2, 3, 5, 7, and 9 T), and S3 (under $H$ = 0, 1, 2, 3, 4, 5, 6, 7, 8, and 9 T), respectively, for \textit{$H$}$\parallel$\textit{$ab$}. (e)--(g) Magnetic field dependence of resistivity of samples S1 (for $T$ = 1.75, 2.45, 3.5, 4.2, 5.5, 6.5, 9, 11.5, 12.5, and 13.5 K), S2 (for $T$ = 1.6, 3.2, 4.2, 6, 8, 11, 12.5, 13.2, 14, and 16 K), and S3 (for $T$ = 2.45, 3.5, 5, 8, 10, 11, 12, 13, 14.5, 15, and 15.5 K), respectively, for \textit{$H$}$\parallel$\textit{$ab$}. (i)--(k) Temperature dependence of resistivity of samples S1, S2, and S3, respectively. The $H$ values are the same as those of panels (b)--(d) but along the $c$-axis. (l)--(n) Magnetic field dependence of the resistivity of samples S1 (for $T$ = 2, 2.6, 4.3, 6, 8, 10, 11, 12, 13, and 15 K), S2 (for $T$ = 2.1, 3.18, 4.2, 6, 7.5, 9, 10.5, 13.5, 14, and 16 K), and S3 (for $T$ = 2.6, 4.2, 6, 8, 10, 11, 12, 13, and 15 K), respectively,  for\textit{$H$}$\parallel$\textit{$c$}.}
	%\end{center}
\end{figure*}

\section{Fitting on $\mu_0$$H$$_{\rm{c2}}$}
\subsection{$\mu_0$$H$$_{\rm{c2}}^c$}

\begin{figure*}
	\includegraphics[width=18pc]{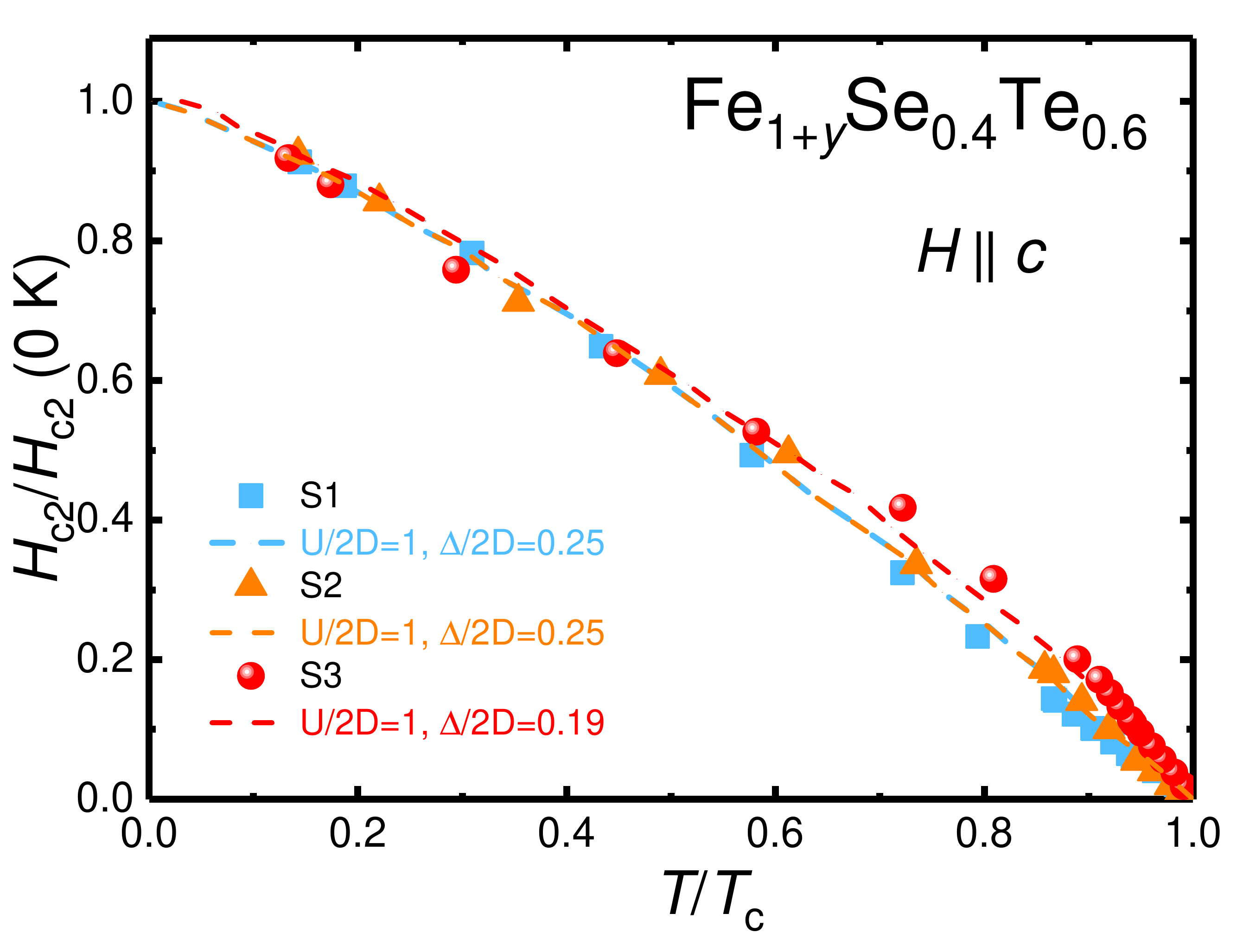}
	%\begin{center}
	\caption{\label{S2}Fit results on $\mu$$_0$\textit{$H$}$_{\rm{c2}}^{c}$ by using BCS-BEC model.}
	%\end{center}
\end{figure*}

The possibility of crossover from BCS to BEC has been discussed in main text. Further investigation on reduced $\mu$$_0$\textit{$H$}$_{\rm{c2}}^{c}$ fitted by Disordered Attractive Hubbard Model model\cite{kuchinskii2017temperature} have been shown in Fig. S2. Here $D$ defines conduction band half-width, $U$ is the Hubbard attraction on the lattice site, distribution width $\triangle$ serves as a measure of disorder, $a$ is the lattice parameter and $b$ is reduced factor. $\triangle$/2$D$ reflects the concentration degree of disorder, $\triangle$/2$D$ $\geq$ 1 is for the dirty limit, while $\triangle$/2$D$ $\leq$ 1 is for the clean limit. $U$/2$D$ reflects the strength of coupling. $U$/2$D$ $\geq$ 1 represents the BEC strong coupling limit, while $U$/2$D$ $\leq$ 1 represents the BCS weak coupling limit. The BCS-BEC crossover happens at  $U$/2$D$ $\sim$ 1. As we can see in Fig. S2, all the three samples exhibit the similar strength of SC coupling with $U$/2$D$ $\sim$ 1, indicating that those samples are located in crossover region from BCS to BEC.

\begin{figure*}
	\includegraphics[width=37pc]{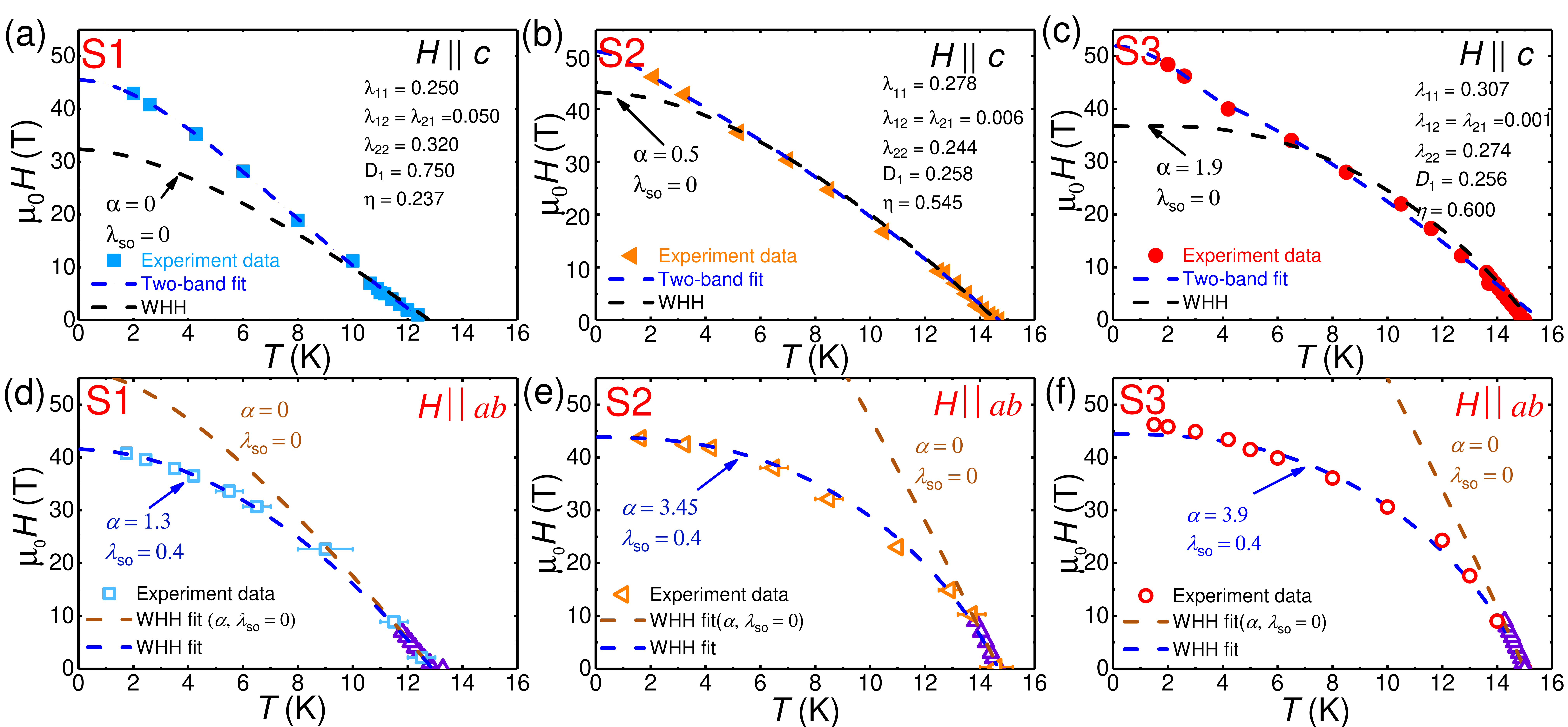}
	%\begin{center}
	\caption{\label{fig3} (a)-(c) $\mu$$_0$$H$$_{\rm{c2}}^c$($T$) of samples (a) S1, (b) S2, and (c) S3 for $H$$\parallel$$c$. The points represent the obtained experimental data, and the blue curves represent the results of the two-band model fitting. Fit results on $\mu$$_0$\textit{$H$}$_{\rm{c2}}^{c}$ of (a) S1, (b) S2, and (c) S3 by using WHH model (black dashed curves) and two-band model (blue dashed curves). $\mu$$_0$\textit{$H$}$_{\rm{c2}}^{ab}$($T$) of samples (d) S1, (e) S2, and (f) S3 for $H$$\parallel$\textit{$ab$}. The blue dashed curves represent the WHH fittings with finite $\alpha$ and $\lambda_{so}$. The brown dashed curves represent the WHH fittings with $\alpha$ = 0 and $\lambda_{so}$ = 0.}
	%\end{center}
\end{figure*}

\begin{figure*}
	\includegraphics[width=36pc]{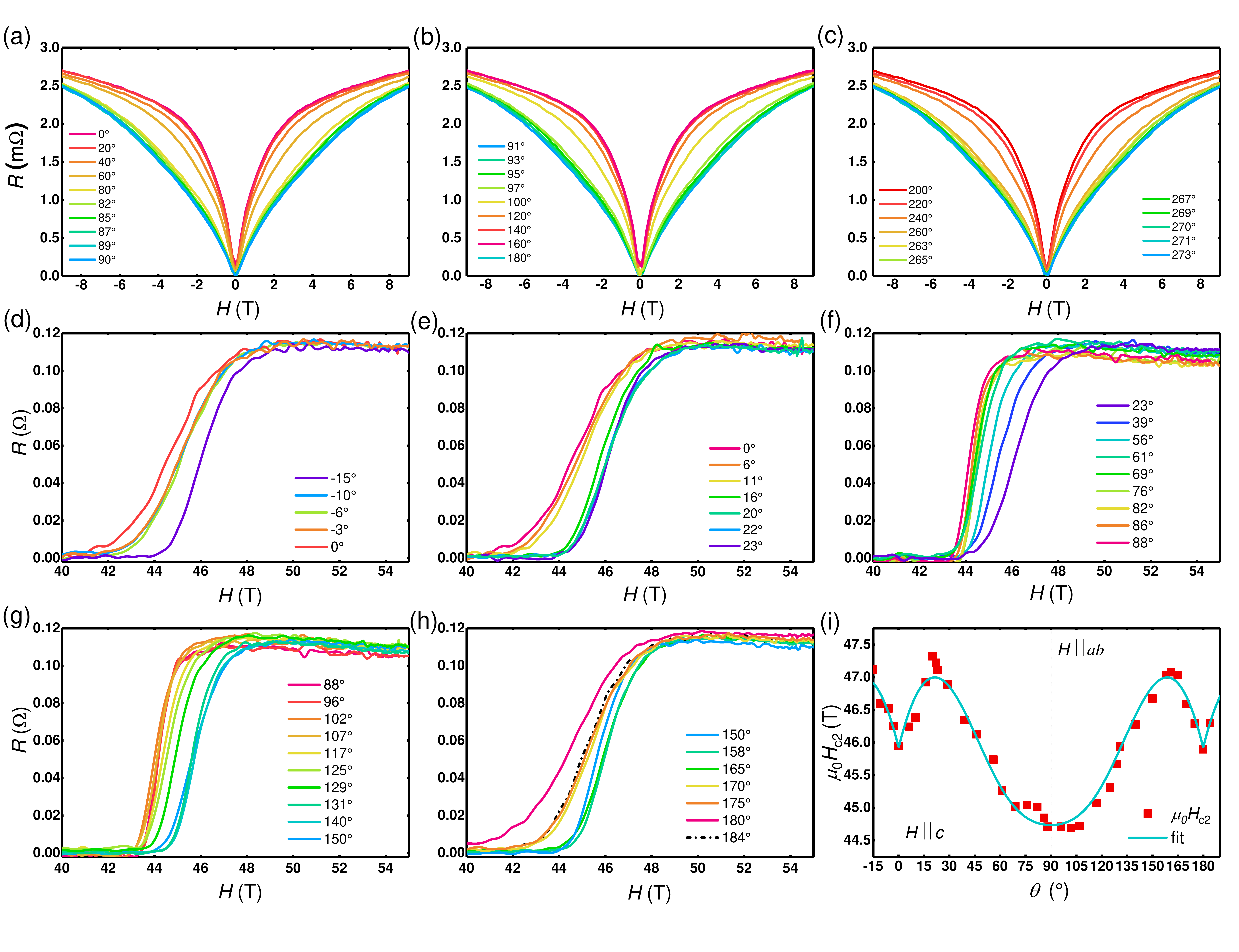}
	%\begin{center}
	\caption{\label{S4} (a)-(c)The $R-H$ curves at $T=$ 14 K with different angle $\theta$ (0$^\circ$ represent $H\|c$ and 90$^\circ$ represent $H\|ab$), which was measured by PPMS. (d)-(h) The $R-H$ curves at $T=$ 2.3 K with different angle $\theta$ (0$^\circ$ represent $H\|c$ and 90$^\circ$ represent $H\|ab$), which was measured in WHHMF. (i) Angle dependent $\mu_0H_{c2}$ at $T=$ 2.3 K, the red points and the azure curve represent the experiment data and the fitting result by equation.(12), respectively.}
	%\end{center}
\end{figure*}

$\mu$$_0$\textit{$H$}$_{\rm{c2}}^{c}$ of S1, S2, and S3 have been plotted in Figs. S3 (a)-(c). Blue dashed curves represent the fitting results by using two-band model as discussed in main text. The parameters also be listed in every figures. Detail discussion can be found in main text. Black dashed curves represent the fitting results by using single-band WHH model (solid curves), which are failed.

\subsection{$\mu_0$$H$$_{\rm{c2}}^{ab}$}
Temperature dependence of $\mu_0H_{\rm{c2}}^{ab}$ of S1, S2, and S3 and the WHH model fitting results are plotted in Figs. S3(d)-(f). The parameters also be listed in every figures. Detail discussion can be found in main text.

\section{Derivation of Extended Anisotropic G-L Model}
According the Ginzburg-Landau (G-L) theory, the Gibbs free energy density of superconductor can be written as following:
\begin{eqnarray}
	&&{g_s}={g_n}+\alpha | \psi | ^2+\frac{\beta}{2} | \psi | ^4+.......
\end{eqnarray}
Here the $\psi$ is the order parameter and $|\psi|^2$=$n_s$, the $n_s$ is the superconducting electron density. The $g_n$ is the Gibbs free energy density in normal state. Due to the $\psi$ should change with spatial position, the equation (1) should be revised as following:
\begin{widetext}
	\begin{eqnarray}
		&&{g_s(H_a)}=f_n(0)+\alpha | \psi | ^2+\frac{\beta}{2} | \psi | ^4+\frac{1}{2m}|-i\hbar \bigtriangledown \psi-eA\psi|^2+ \frac{B^2}{2\mu_0}-B*H_a.
	\end{eqnarray}
\end{widetext}
The angle-dependence of Zeeman splitting energy and orbital pair-breaking energy should be considered (Fig. S5). As shown in Fig.S5, Zeeman splitting energy can be written as  $\bigtriangleup E_z$ = 2 $|M\cdot H|$ = $|2 g_{ab}\mu_BH\sin\theta |$. Since the Zeeman splitting energy does not depend on the sign of $\sin\theta$, therefore, the absolute value has been added to make the $\bigtriangleup E_z$ an even function with positive values when $\theta$ is away from [$0^\circ, 180^\circ$]. Orbital pair-breaking energy can be written as $\bigtriangleup E_{\rm{orb}} =|2F_L\cdot  \xi_{GL}(\theta) /2|= | eHv_{ab}|\xi_{GL}(\theta)$, where the $F_L$ is the Lorentz force. The Lorentz force $F_L$ shows a maximum value only when the orbital planes of Cooper pairs are perpendicular to $H$. Corresponding, $v$ is always parallel to $ab$-plane ($v_{ab}$) and perpendicular to both $H$ and $F_L$.

\begin{figure*}
	\includegraphics[width=34pc]{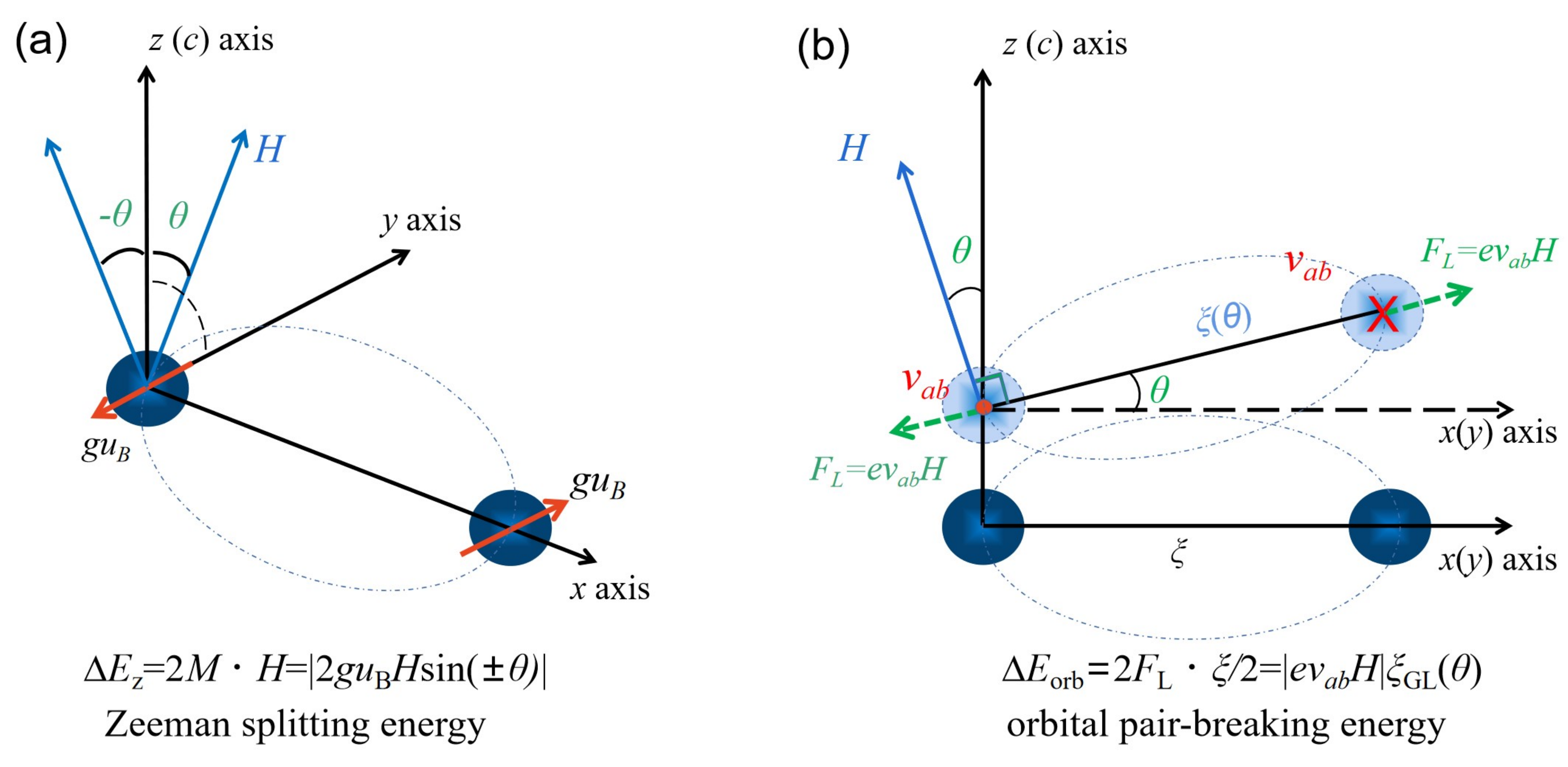}
	%\begin{center}
	\caption{\label{fig5} Schematic diagrams of (a) Zeeman splitting energy and (b) orbital pair-breaking energy. }
	%\end{center}
\end{figure*}

Now we add the Zeeman splitting energy and orbital pair-breaking energy into equation (2)\cite{fuchs2009orbital}:
\begin{widetext}
	\begin{eqnarray}
		&&{g_s(H_a)}=f_n(0)+\alpha | \psi | ^2+\frac{\beta}{2} | \psi | ^4+\frac{1}{2m}|-i\hbar \bigtriangledown \psi-eA\psi|^2+ \frac{B^2}{2\mu_0}-B*H_a+|2g_{ab}\mu_B H\sin\theta |+ |eHv_{ab}|\xi_{GL}(\theta).
	\end{eqnarray}
\end{widetext}
$g_{ab}$ and $v_{ab}$ represent the Lande $g$-factor and Fermi velocity in $ab$-plane, respectively. When the system is stable, $\psi$ adjusts itself to minimize the overall free energy, this variational problem leads to the celebrated G-L differential equation\cite{tinkham1996introduction}:
\begin{eqnarray}
	&&\frac{1}{2m}|-i\hbar \bigtriangledown -eA|^2 \psi+ \alpha \psi +\beta |\psi|^2 \psi =0.
\end{eqnarray}
This G-L differential equation is suitable for both equations (2) and (3). Combining the definition of $\xi_{\rm{GL}}$:
\begin{eqnarray}
	&&\xi_{GL} ^2 (T)=\frac{\hbar ^2}{2m^*|\alpha (T)|}.
\end{eqnarray}
Equation (4) can be written as:
\begin{eqnarray}
	&&{ (\frac{\nabla}{i}-\frac{2 \pi A}{\Phi _0}) }^2 \psi = \frac{2m^*\alpha}{\hbar ^2}\psi = \frac{\psi }{\xi_{GL} ^2(T)} .
\end{eqnarray}
Firstly, when $H||c$-axis, $A_y=Hx$, Equation (6) can be written as:
\begin{eqnarray}
	&&( -\nabla ^2+\frac{4i \pi }{\Phi _0} Hx \frac{\partial }{\partial y} + (\frac{2 \pi H }{\Phi _0})^2 x^2) \psi = \frac{\psi }{\xi_{GL} ^2} .
\end{eqnarray}
Since the effective potential depends only on $x$, it is reasonable to look for a solution of the form, $\psi (x,y,z)=e^{ik_y y} e^{ik_z z} f(x)$, substituting this into equation (7), we find:
\begin{eqnarray}
	&&-f''(x)+(\frac{2 \pi H }{\Phi _0})^2 (x-x_0)^2 f= (\frac{1}{\xi_{GL}^2}-k_z ^2)f .
\end{eqnarray}
where $x_0=(k_y \Phi_0)/2\pi H$. According the equation (8), The total de-pairing energy $\Delta E$ should be equated to $[\hbar ^2 (\xi_{\rm{GL}} ^{-2} -k_z ^2 )]$/$2m^*$:
\begin{widetext}
	\begin{eqnarray}
		&&\Delta E=\Delta E_n +\Delta E_z+ \Delta E_{orb}\nonumber\\
		&&= (n+\frac{1}{2})\hbar \frac{2eH}{m^*c}+ | 2g_{ab}\mu_B H\sin\theta |+ |eHv_{ab}|\xi_{GL}(\theta)=\frac{\hbar ^2 (\xi_{GL} ^{-2} -k_z ^2 )}{2m^*}.
	\end{eqnarray}
\end{widetext}
The $H$ has its highest value ($H_{c2}$) if $n$ = 0 and $k_z$ = 0, so we get\cite{braithwaite2010evidence,klein2010thermodynamic}:
\begin{widetext}
	\begin{eqnarray}
		&&H_{c2}=\frac{\hbar^2}{2m^*\xi_{GL} ^2(\theta)[\frac{\hbar e}{m^*c}+|2g_{ab}\mu_B \sin\theta | +|ev_{ab}|\xi_{GL}(\theta)]} .
	\end{eqnarray}
\end{widetext}

Here we take
\begin{eqnarray}
	&&\xi_{GL} ^2(\theta)=\frac{(\xi_{GL}^{ab})^2}{\sin^2 \theta+\gamma_{GL}^2 \cos ^2 \theta} .
\end{eqnarray}
into equation (10), we can get\cite{sun2019review,stoner1945xcvii,homes2011optical}:
	\begin{widetext}
		\begin{eqnarray}
			&&H_{c2}(\theta)=\frac{\hbar^2}{2m^*\frac{(\xi_{GL}^{ab})^2}{\sin^2 \theta+\gamma_{GL}^2 \cos ^2 \theta}[\frac{\hbar e}{m^*c}+| 2g_{ab}\mu_B \sin\theta | +|eV_{ab}| (\frac{(\xi_{GL}^{ab})^2}{\sin^2 \theta+\gamma_{GL}^2 \cos ^2 \theta})^{0.5}]}.
		\end{eqnarray}
	\end{widetext}

When $\theta =0 ^\circ (180 ^\circ )$, there only angle-dependent orbital pair-breaking effect is active. When $\theta \neq 0 ^\circ (180 ^\circ )$, both angle-dependent orbital pair-breaking effect and Zeeman splitting effect are active.

Angle dependence of $\mu_0H_{c2}$(2.3 K) and its fitting by equation (12) are shown in Fig. S4(i). The polar diagram of $\mu_0H_{c2}$(14 K) and $\mu_0H_{c2}$(2.3 K) can be found in Fig. 4(a) and Fig. 4(b) in main text. Detail measurement data are shown in Figs. S4(a)-(c) with PPMS and Figs. S4(d)-(h) with WHHMF.

\section{Anisotropy of $\xi$($T$)}
The temperature dependence of $\xi$($T$) is discussed for sample S3 without the influence from excess Fe. $\xi$($T$) for the single-band G-L model using the following relationship:
\begin{eqnarray}
	&& \xi(T) = \xi(0 K)(1-T/T_c)^{-1/2},
\end{eqnarray}
which has been applied to FeSe\cite{zhou2021disorder} and Ca-112 system\cite{xing2017two}. On the other hand, for the two-band model, it follows the relationship\cite{askerzade2002ginzburg}:
\begin{widetext}
\begin{eqnarray}
	&& \xi(T) = \xi(0 K)[f_1(1-T/T_c)+f_2(1-T/T_c)^2]^{-1/2},
\end{eqnarray}
\end{widetext}
where $f_1$ and $f_2$ are coefficients. This equation has been used for MgB$_2$ and LuNi$_2$B$_2$C\cite{grigorishin2016effective,askerzade2002ginzburg}.
\begin{figure}
	\includegraphics[width=18pc]{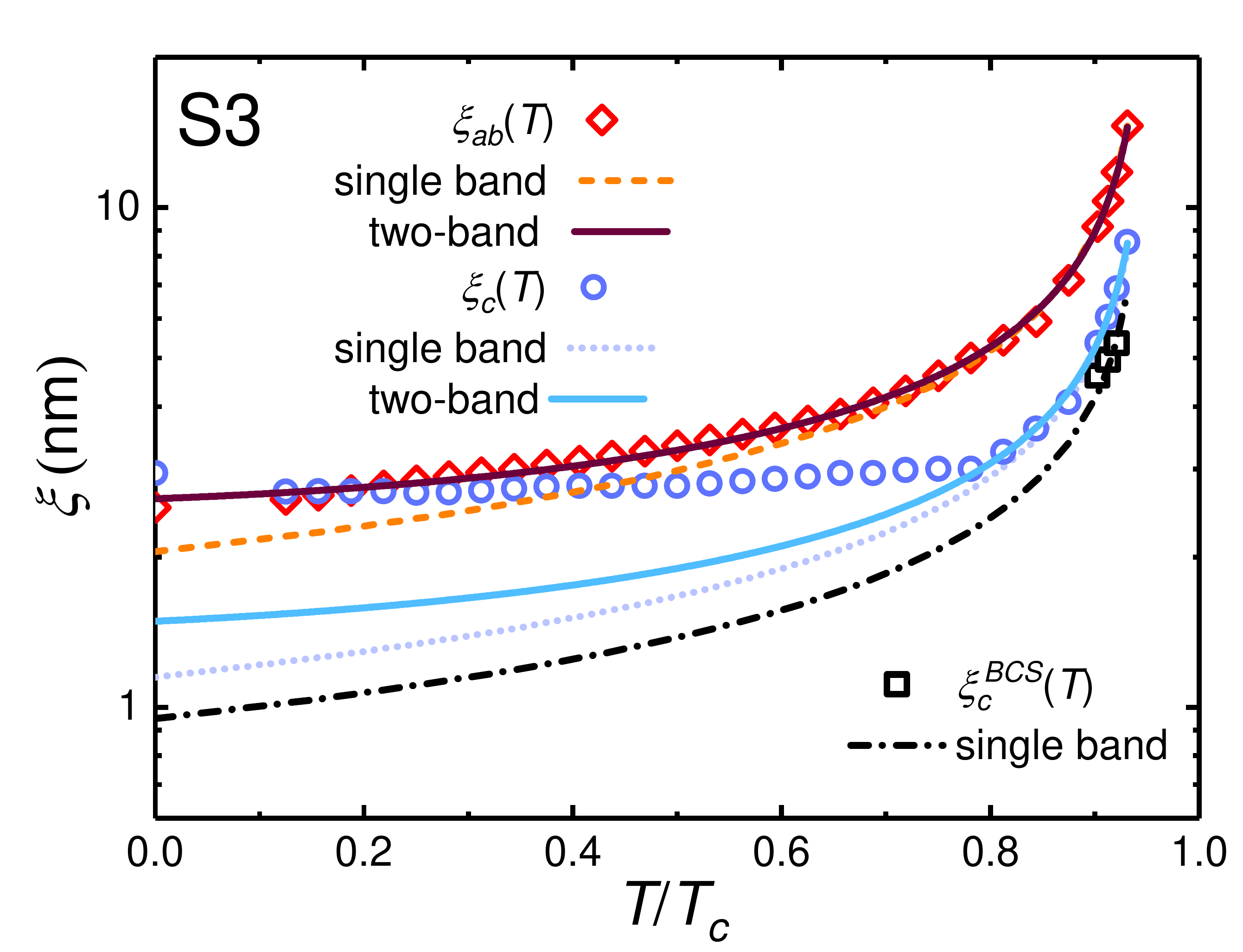}
	%\begin{center}
	\caption{\label{fig6}Temperature dependence of $\xi$($T$). The red and blue points represent $\xi$$(T)$ in two directions. The orange- and blue dashed curves represent single-band model fittings for the two directions, whereas the brown and blue solid curves represent the two-band model fittings results for the two directions. The black points and dashed curve represent $\xi_c^{\rm{BCS}}$($T$) and its fitting obtained using the single-band model, respectively.}
	%\end{center}
\end{figure}

When $H$$\parallel$$c$, since there is no spin-paramagnetic effect in this direction, $\xi$$_{ab}$($T$) (red diamonds in Fig. S6) with 0 K $\leq$ $T$ $\leq$ $T_c$ can be obtained from ($\Phi$$_0$/2$\pi$$\mu$$_0$$H$$_{\rm{c2}}^c$($T$))$^{1/2}$. As shown in Fig. S6, the fitting for $\xi$$_{ab}$($T$) obtained using the single-band model (orange dashed curves) deviates from the experimental data when $T$ $<$ 0.7$T_c$. Contrarily, the fitting obtained using the two-band model (brown dashed curves) reproduces the experimental data very well over the whole temperature region. It is consistent with the two-band model fitting for $\mu$$_0$$H$$_{\rm{c2}}^c$ (Fig. S3(c)), proving that Fe$_{1+y}$Te$_{0.6}$Se$_{0.4}$ is a multigap system.

A peculiar situation occurs for $\xi_c$($T$) obtained from $\Phi$$_0$/2$\pi$$\xi$$_{ab}$($T$)$\mu$$_0$$H$$_{\rm{c2}}^{ab}$($T$). $\xi_c$($T$) and the fittings obtained using equations (13) and (14) are plotted in Fig. S6 with blue circles, blue dots, and an azure dashed curve, respectively. Clearly, both fittings  deviate strongly from the experimental data except for the region close to $T_c$. This poor fitting is also typical of many other superconductors\cite{askerzade2002ginzburg,vedeneev2006reaching}. This demonstrates that both the single-band and two-band theories fail to explain the $\xi$$_{c}$($T$) trend, which is due to the strong spin-paramagnetic effect, as discussed above.

$\xi_c$(0 K) should be obtained without the influence of the spin-paramagnetic effect. According to the two-band BCS theory,
when $T$ $\rightarrow$ $T_c$, $\xi$$_{c}^{\rm{BCS}}$($T$) can be expressed as follows\cite{gurevich2011iron}:
\begin{widetext}
\begin{eqnarray}
	&& \xi_{c}^{\rm{BCS}}(T) = \Phi_0/(2 \pi T_c(-\mu_0dH_{c2}^{ab}(0 \rm{K})/dT)\xi_{\textit{ab}}(T)).
\end{eqnarray}
\end{widetext}
Due to the single-band behavior of $\mu$$_0$$H$$_{\rm{c2}}^{ab}$, $\xi_{c}^{\rm{BCS}}$($T$) was fitted using equation (15). Both $\xi_{c}^{\rm{BCS}}$($T$) near $T_c$ (black hollow blocks) and the fitting result (black dashed curve) are plotted in Fig. S6. $\xi_{c}^{\rm{BCS}}$(0 K) was roughly estimated as 9.5 \AA, which is slightly larger than previous report $\sim$ 4.4 \AA \cite{gurevich2011iron}. The anisotropy of $\gamma_{\xi}$(0 K), defined as $\xi_{ab}$(0 K)/$\xi_{c}^{\rm{BCS}}$(0 K), is 2.64 and is consistent with ($m_c$/$m_{ab}$)$^{1/2}$ ($\sim$ 2.5 -- 4)\cite{sun2021comparative}, indicating that the superconductivity is anisotropic. Compared with the lattice parameter $c$ $\sim$ 6 \AA, this small $\xi_{c}^{\rm{BCS}}$(0 K) indicates that Fe$_{1+y}$Te$_{0.6}$Se$_{0.4}$ may be a quasi-two-dimensional superconductor.

\end{document}